\documentclass{article}
\usepackage[dvips]{graphicx}
\begin{document}

\pagestyle{plain} 
\setcounter{page}{1}
\setlength{\textheight}{725pt}
\setlength{\topmargin}{-40pt}
\setlength{\headheight}{0pt}
\setlength{\marginparwidth}{-10pt}
\setlength{\textwidth}{20cm}

\title{Network Growth via Preferential Attachment based on Prisoner's Dilemma Game }
\author{Norihito Toyota \and Hokkaido Information University, Ebetsu, Nisinopporo 59-2, Japan \and email :toyota@do-johodai.ac.jp }
\date{}
\maketitle
\begin{abstract}
In this article we discuss network growth based on Prisoner's Dilemma Game(PDG) where palyers on nodes in a network 
palay with its linked players. 
The players estimate total profits according to the payoff matrix of the PDG. 
When a new node is attached to the network, the node make linkes to nodes in the network 
with the probabilities in proportion to the profits made by the game. 
Iterating this process, a network grows. 
We investigate properties of this type of growing networks, especially the degree distribution and 
time-depending strategy distribution  by running  computer simulation. 
We also find a sort of phase transition in the strategy distributions.  
For these phenomena given by computer simulation, theoretical studies are also carried out.   

\end{abstract}
\begin{flushleft}
\textbf{keywords:}
Network Growth, Game Theory, Prisoner's Dilemma, Degree Distribution, Phase Transition
\end{flushleft}

\section{Introduction}
At the close of the 20th century, many empirical networks in real world turn out to be  scale free or 
 small world networks\cite{Albe1,Albe2,Albe3,Watt1,Watt2}, rather than random networks\cite{Erdos1}. 
At the same time, it becomes clear that in what way should we construct these networks. 
In those ways networks are mainly constructed by some outside algorithms, 
only depending of the topology of the networks\cite{Newm}.   
In the growing way such as preferential attachment\cite{Albe1,Albe2}, priferential nodes will 
forever preserve the position as preferrer. 
In real systems, however, such situations are not universal.   
Depending on a dynamics of a system, priferential nodes may change moment by moment. 
In this article we consider the innner dynamics on a network to influence which are preferential nodes.  
Thus we explore the interactions between an inner dynamics on a network and network growing.  
Such network growing seems to be more universal than models without some inner dynamics. 
Though many flexible model have been proposed, in which fitness\cite{Bian}, aging\cite{Klem1,Klem2,Doro1,Doro2}
, hierarchy\cite{Rava} and so on have been considered, 
there are little dynamical models accompanied with interactions among nodes in a network.  

As considering an innner dynamics on a networks, we need to introduce some inner degree of freedom  
to realize interactions. 
In this article we adopt Prisoner's Dilemma game (PD)\cite{Axel1,Poud1,Rapo2}, 
which is used in various fields, economics, sociology, biology and so on,  as a dynamics on  networks. 
In this case, the types of strategies, cooperation (C) and defection (D), 
which correspond to the inner degree of freedom are two. 
Iteration of games played by players arranged in two dimensional space has been first considered 
on regular lattice by Nowark and May\cite{Nowa1,Nowa2}.   
After that many reserchers have studied the subject\cite{Weib1}. 
Researchs in game theory related to complex networks, which was anticipated by Ref. \cite{Abra1}, are summarized 
 in Ref. \cite{Szab3} well.  
The relations between a network topology and a game dynamics on the network have been also investigated by some athors
\cite{Zimm1,Zimm2,Zimm3,Zimm4,Holm1,Biel1,Ebel1,Ebel2,Kim1,Toyo7}. 

In this article we use game thaory for growing networks\cite{Toyo8}.   
We first simulate network growing based on PDG dynamics by changeng a parameter that represents payoff of PDG, 
and analyze the results. 
We first focus the time series of the population of two strategies.  
A phase transition like phenomenon appears when changing the parameter where all D world turns into all C world all at once.   
Then we investigae the degree distributions of resultant networks.  
They are mainly classified into two types of forms, depending on the parameter. 
The turnning point is the critical point in the D-C transition.  
We also discuss these phenomena from theoretical point of view and some are quantitatively derived.    
A comment is also made on a relation between a sahpe of probability function and degree distribution function. 

In the section two we formulate the algorithm for network growing studied in this article.  
We give simulation results in the section three and theoretical analyses are given in the next section. 
The last section, five, is devoted to summary.  

\section{Model for Network Growing}
We describe the model for  network growing investigated in this article. \\

1. Start a complete graph with $k+1$ nodes on which  either strategy C or D are assigned at \hspace*{8mm} random.  \\

2, The nodes play PDG with their (first $k+1$) linked nodes including themselves on the network\\

3. Each node $i$ evaluates the total payoff $R_i$ acquired from her/his neighborhood according to 
 \hspace*{8mm} the payoff matrix given by Table 1.   \\

4. Each node $i$ mimics the strategy of the node with the highest payoff among linked nodes. \\

5. Add a new node with a strategy chosen randomly. \\

6. Then new links from the new node  to  $k$ ones among already existing nodes 
are connected  \hspace*{8mm} with the probability $P_i$ depending on the total payoffs of the exsiting nodes;  
\begin{equation}
P_i = \frac{1}{1+\exp(-R_i / A)},  
\end{equation}
 \hspace*{8mm} where $A$ is some positive constant. \\

7. The procedures from  2. to 6. are iteratively carried out $n$ times. \\

As considering preferential attachments relative to payoff, it is natural that the probability $P_i$ 
that are linked with a newly joining node is proportional to acquired payoffs. 
Now notice that payoffs can take both positive and negative values, since total payoff is zero in the payoff matrix of PDG.  
Then we use Eq. (1) as an extention of simple preferential attachment in the step 6. 
A little discussion will be given for this later on.  
We mainly analyze the time series of population of nodes with C or D strategy and degree distribution 
by commputer simulations.   
\begin{center}
Table 1.  Payoff table(matrix) of PD game on which $t>c>d>S$ is imposed.  \\
\begin{tabular}{|l|c|c|} \hline 
strategy   &C  &D \\ \hline \hline
C  & ($c$, $c$) &  ($s$, $t$) \\ \hline
D  & ($t$, $s$) &  ($d$, $d$) \\ \hline
\end{tabular}
\end{center}

\section{Results of Computer Simulations}
We choose $k=5$ , $A=30$ and $t=-s=5$ as parameters 
for the procedure for  network growing described in the previous, 
and $n=1000$ as long as convergence is out of question.  
We determine $c+ d + s +t=0$ so that the average payoff can be 0 in the payoff matrix. 
So $c=-d$ is a variable.  
It is studied how the features of a network system vary according to this variable $c$.

\subsection{Degree Distribution}
Main results of simulations are the following. \\
・For $c<4$, the strategy of the network converges into all D and then degree distribution looks like linear (See Fig.1).\\
・For $c>4.4$, the strategy of the network converges into all C and then degree distribution looks like exponential one 
(See Fig.2).

\begin{center}
\includegraphics[scale=1.0,clip]{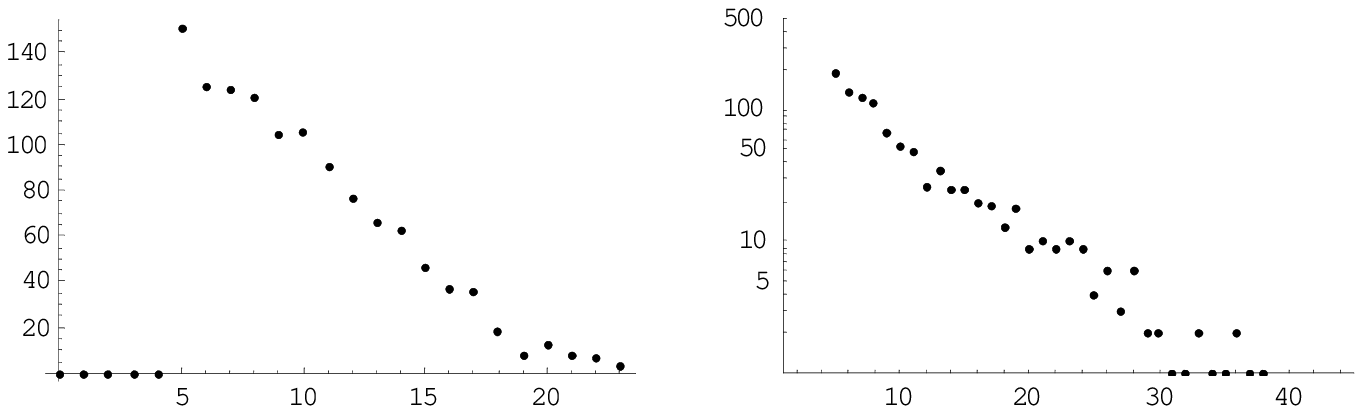}\\
Fig1. Degree distribution for n =1200, c =4.0　
Fig2. Degree distribution for n =1000, c = 4.5 \\
\hspace{8cm} in a logarithmic scale for the vertical axis. 
\end{center}　

 According to one's expectation, D as a dominant strategy  is prevalent as $c$ takes small values. 
As $c$ gets bigger, all nodes comes to take a strategy C. 
When the system shifts from all D to all C world at a critical $c$ value, the feature of degree distribution changes as well. 
 The abovementioned features  in degree distribution is what is presumed phenomenologically. 
The system becomes unstable in the middle value of $c$ where stochastic behavior appears, that is, 
all C on one occasion, all D on another occasion. 
The ratio of cases where  the strategy of  systems becomes  all C is shown in Fig.3. 
While this is an average over about 10 times iterations, it is clear that a sort of phase transition occurs with 
a order parameter $c$. 
D, however,  is prevalent in the early of simulations in all cases. 
At $c> 4$, the cases where C drastically becomes prevalent with advancing time steps take palce, which is shown in the Fig 4. 
We will make theoretical anayses of these in the next section.  
    
\begin{center}
\includegraphics[scale=0.9,clip]{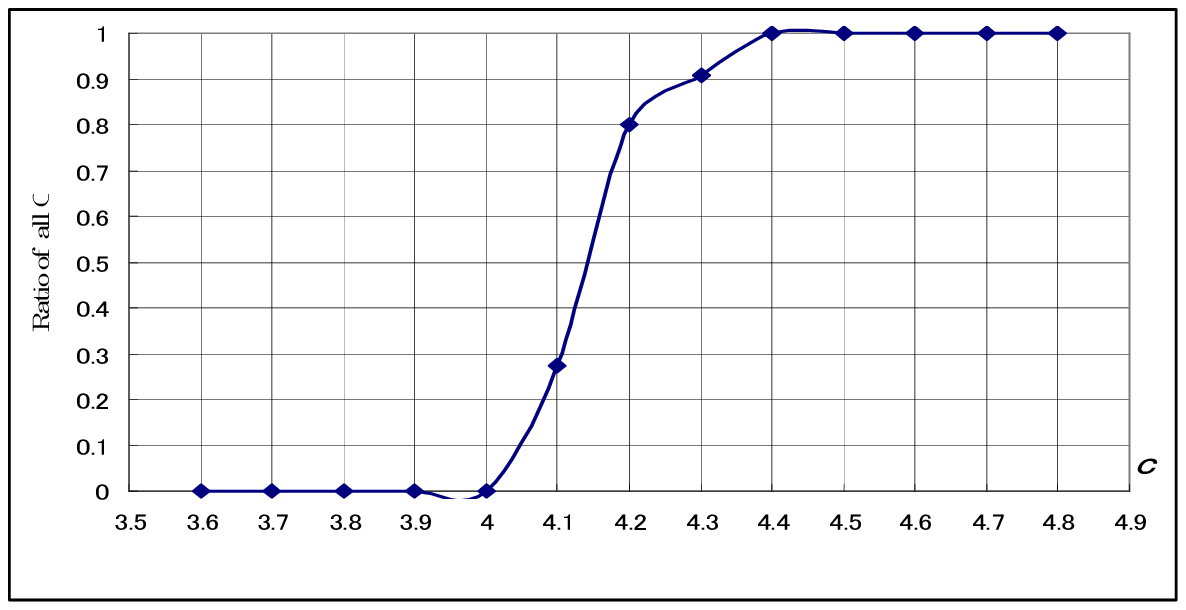}\\
Fig. 3 $c$ v.s. probability of all C. 
\end{center}　

\section{Theoretical Analyses}
\subsection{D-C Transition}  
Even if the strategy of a new node added to the system  is C, the strategy of the node turns from C to D 
when the total payoff of the newly joining node is not greatest among all nodes connected to the newly joining node.   
When C joins in a network system with all D, the payoff $P_c$ that C gets is given by 
\begin{equation}
P_c=c + ks=c-tk.
\end{equation}

The payoff $P_d$ of nodes with D connected to the newly joining node is given by 
\begin{equation}
P_d=t+xd=t-xc,
\end{equation}
where $x$ is a degree of a node with strategy D. 

By demanding $P_c > P_d$, the condition that the strategy C of the newly joining node does not turun to D leads to  
\begin{equation}
c+ks >t+xd.
\end{equation}
After all we obtain a condition
\begin{equation}
c > \frac{1+k}{1+x} t  \; \mbox{ with } k=t=5.  
\end{equation}

Minimal values of $x$ which satisfy the inequality (5) for various $c$ values are given in Table 2. 
The situation changes at nearly the critical value of $c$ pointed out in the previous section. 
Considering that $x$ is originall meaningful only when $x$ is interger, because $x$ is a degree, 
it leads to the fact that  while for $c>4.3$ a newly joining C node can change D nodes with $x=6$ to C nodes when the D nodes are connect 
with the C nodes, for $c<4.3$ it can do D nodes with only  $x=7$.  
Though the boundary value of $x$ is yet a magic number, we find that there is a gap between regions with $c>4.3$ and $c<4.3$ 
how a newly joined strategy C can easily make a strategy D change into C.   

$ c>2$ is also enough for this C to propagate through D-dominant world.  
As a natural conjecture, it is considered that the magic number $x=6$ has some significance and 
there may be a gap in  strategy distribution at a point near $c=4.3$.    
This fact may be also  supported by Fig. 3.  
We pursue this conjecture. 
\begin{center}
Table 2. Minimal values of $x$ satisfying the inequality (5). \\
\begin{tabular}{l|c|c|c|c|c|c|c|c|c|c|c|c|c} \hline 
c   &2  &3 &3.9 &4.0 &4.1 &4.2 &4.3 &4.4 &4.5 &4.6 &4.7 &4.8 &4.9 \\ \hline \hline
x  &14 &9 &6.7 &6.5 &6.3 &6.1 &6.0 &5.8 &5.7 &5.5 &5.4 &5.3 &5.1\\ \hline
\end{tabular}
\end{center}

\begin{center}
\includegraphics[scale=0.8,clip]{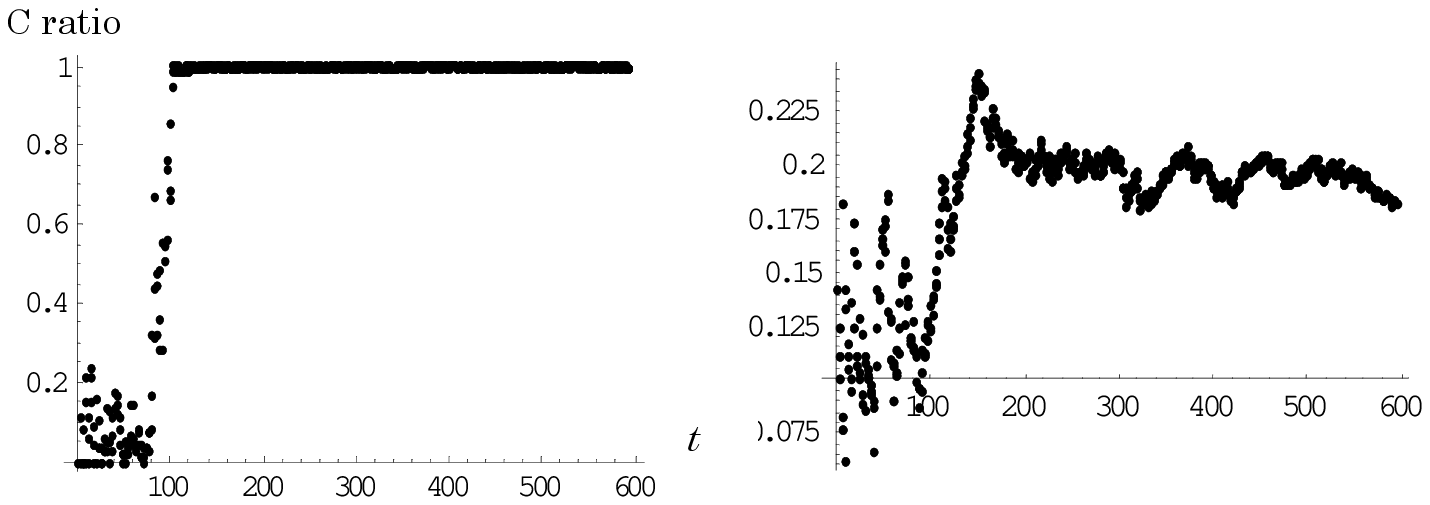}\\
Fig. 4 D-C transition and the ratio of degree$=5$ among all nodes in $n=600$ and $c=4.3$. 
\end{center}　

\begin{center}
\includegraphics[scale=1,clip]{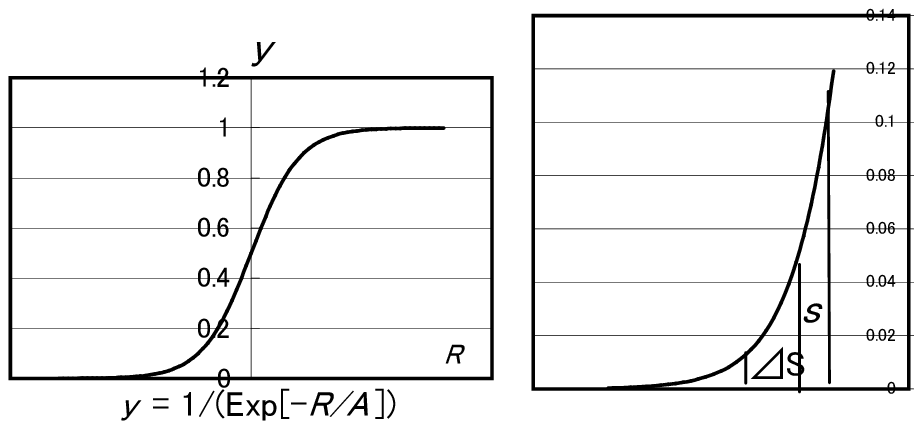}\\
Fig. 5 Probability function and its enlarged drawing.  
\end{center}　

 First of all, Fig. 4 gives a suggestion to study the conjecture.  
The ratio of nodes with low degree ($x<6$)  drastically decreases in the right-hand side of Fig.4 temporarily 
just when all D world turns to all C world. 
 After that the ratio increases almost smoothly with time steps. 
A decrease in the population of nodes with low degree induce nodes with the strategy D to the strategy C,
 acoording the inequality condition (5). 
The reason is that a newly joining node is apt to mainly builds linkes with nodes with highly degree due to the decrease.    

We consider the condition that a newly joining node connects with nodes with high degrees on a network 
with a probabolity of not less than 50 percent. 
From that, we show that critical value of $c$ in the C-D transition can be estimated. 
This is  due to the fact that the probability that a node links with a newly joined node depends on total payoff $R_i$, 
which also depends on $c$.    
Let us introduce  a possibility $p$  that a node has not less than degree $x$.
From the condition that all five ($k$ in general) nodes linked with a newly joining node with strategy C have 
not less than  $x$  degrees, 
we obtain a constraint on $p$; 
\begin{equation}
p^5>0.5 \longrightarrow p>0.87.
\end{equation}

In the meanwhile, degree distribution for all D world is a linear function, which has phenomenologically 
  a general form  (the leftt-hand side of Fig. 1)
\begin{equation}
Y=\alpha R +\beta\;\mbox{ with }  \; 
\alpha = -\frac{2n}{15^2c}\;<<\;\beta=\frac{40n}{15^2},\;\;R=kc.
\end{equation}
Nodes with not less than $k=9.4$ degree have a majority in all. 
This value is also an average one for degree of all nodes which agrees with simulation results 
not explicitly demonstrated in this article.  

In order to specify the condition that a newly joining node with strategy C  can link with nodes with high degree 
with a high probability, we first estimate the normalization factor of the probability in Eq. (1). 
The total sum of $P_i$ over all nodes is 

\begin{eqnarray}
S(d)&=& \sum_{i=1}^n \frac{\alpha R_i +\beta}{1+e^{-R_i/A}} \sim \int_{R_{min}}^{R_{max}} \frac{\alpha R +\beta}{1+e^{-R_i/A}}dR 
\sim \int_{6d}^{21d} \frac{\beta}{1+e^{-R_i/A}}dR \nonumber\\
&=& \beta A \left( \ln [e^{6d/A} +1] - \ln [e^{21d/A} +1] \right).
\end{eqnarray}
By using Eq. (7), the sum over nodes is changed to the sum over payoffs in the first equality. 
Since the sum of the above sequence can not be analitically calculated, a continuous approximation is maded in Eq. (8). 
Furthermore using the inequality in Eq. (7), a analytic expression for $S(d)$ is obtained in the last equality.    
If the partial set  $\Delta S$ of the gross area noted by $S$ in Fig. 5 satisfys the following inequalty
\begin{equation}
\frac{\Delta S(d,x)}{S(d)}> 0.87,
\end{equation}
degree of nodes connected to a newly joining node with C would be larger than $x$ in all likelihood.  
Here the explicit representation of the fraction $\Delta S$ is given by
\begin{eqnarray}
\Delta S(d,x) &\equiv&  \int^{d(1+x)}_{21d} \frac{\beta}{1+e^{-R_i/A}} =
 \beta A \left( \ln [e^{d(1+x)/A} +1] - \ln [e^{21d/A} +1] \right).
\end{eqnarray}


Finally  we can derive a critical velue $d$ by soluving the following  equation;  
\begin{equation}
 \frac{\Delta S(d,x=k_c)}{S(d)}= 0.87,   
\end{equation}

Though this equation has two unknown quantities $k_c$ and $d$,  
we can be solved the equation to find the value $d$ numerically;
\begin{equation}
\left\{ 
\begin{array}{l}
 d=0.29 \; \mbox{ for } x=7, \;  \mbox{ this is nonsense},   \\
  d= -4.14 \; \mbox{ for } x=6  . 
\end{array}\right.
\end{equation}

Note that the solution $d=0.29$ for degree $x=7$ is meaninglessness,  but  
the meaning solution $d=-4.14$, which accords with Fig.3  well, at last appears for $x=6$,  that is a just magic number.   


\subsection{Degree Distribution Function}
We follow the method adopted by Brabashi et al.\cite{Albe2} in order to find a degree distribution function 
in this model. 
The corresponding equation to the one used by Brabashi et. al.\cite{Albe2}  for $k_i$ in the present model is 
given by 
\begin{equation}
\frac{\partial k_i(t)}{\partial t}= \frac{m}{S(1+e^{-R_i/A})}=\frac{m}{S(1+e^{-r k_i/A})},  
\end{equation}
where $r=c \;\;or\;\; d.$
$S$ at a time step $t$ for both cases, all C and all D, is a linear function of $t$ as shown in Fig.6; 
\begin{equation}
 S \sim \gamma t.
\end{equation}
where $\gamma$ is some proportionality. 

Solving the differential Eq. (13) under the initial condition of  $t=t_0,\; k_i=m=5$, 
we obtain 

\begin{equation}
(k_i(t)-m)+\frac{A}{r} (e^{-rm/A}- e^{-rk_i(t)/A}) = \frac{m}{\gamma} \log_e \frac{t}{t_0}. 
\end{equation}

Though we can not give any analytic expression of $k_i (t) $, we have only to find a $t_0$ derivative of $k_i (t) $. 
According to Barabashi et. al.\cite{Albe2}, a distributon function can be given by 
 \begin{equation}
P(k)=-\frac{1}{ \partial k_i(t) / \partial t_0}. 
\end{equation}
So we obtain the below equation;
\begin{eqnarray}
P(k)& \sim&  \frac{\gamma t}{5} \frac{1+e^{-r k/30}}{e^{\gamma(k-5)/5} e^{6\gamma/r}(e^{-\gamma/6} -e^{-rk/30})},  
         \nonumber \\
    &=& \frac{\gamma t}{5} \frac{\coth{r k/60}}{e^{\gamma(k-5)/5} e^{6\gamma/r}}\\
 &\sim& O(1)- O(k^1) \;\;\mbox{for  $r=d$}\\
&\sim&  e^{\gamma(5-k)/5}  [O(1)- O(k^1) ]  \;\;\mbox{for  $r=c$}
\end{eqnarray}
 Eq. (18) in the third line shows that  $P(k)$  linealy decrease with $k$ approximately in all D world and 
Eq. (19)   roughly shows an exponential damping with respect to $k$ with some correction at  in all C world. 
Both of them are consisyent with simulation results in section2. 
In fact both theoretical estimation coming from Eq. (17) and simulation results are comapared 
for an all D case and an all C case  in Fig.8 and 9, respectively.  
The results of simulations conform with theoretical ones for both cases well. 
 
\begin{center}
\includegraphics[scale=0.7,clip]{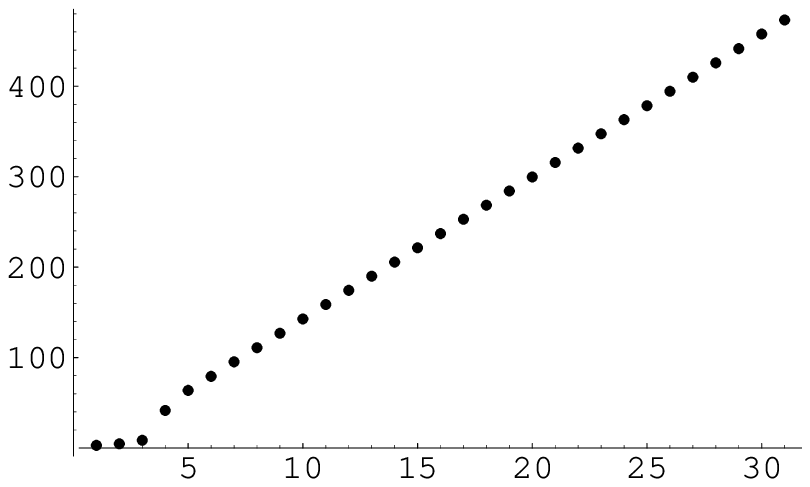}
\includegraphics[scale=0.7,clip]{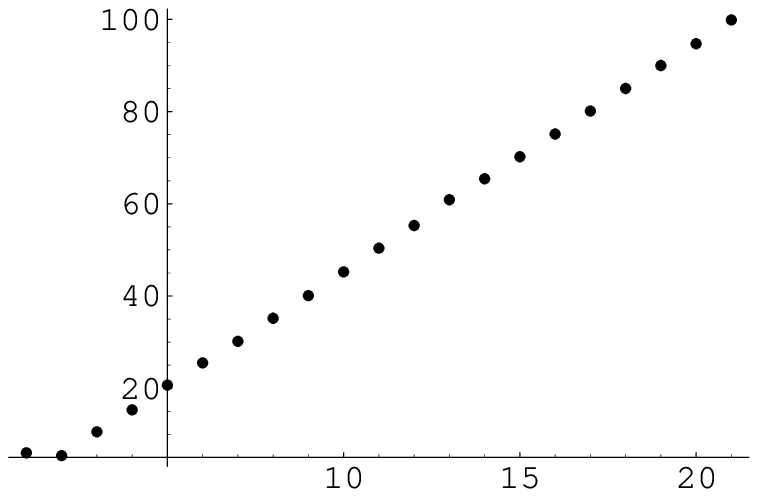}\\
Fig.6 $S(t)$ for all C world (left) and all D world (right). 
\end{center}

\begin{center}
\includegraphics[scale=0.8,clip]{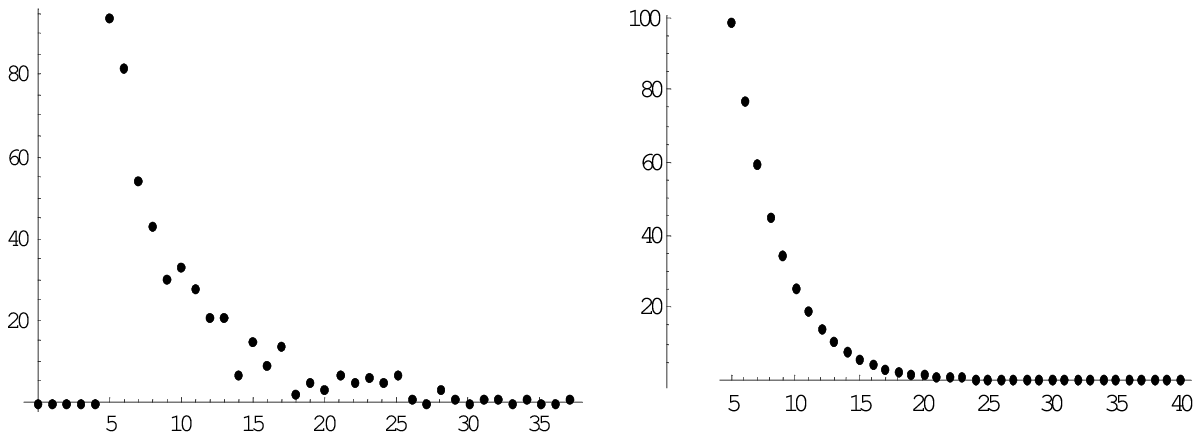}\\
Fig.7 Degree distributions of simulation for $n=500$ and $k=5$ (left) and theory ($\gamma = 1/3$) (right) at $c=4.5$

\includegraphics[scale=0.8,clip]{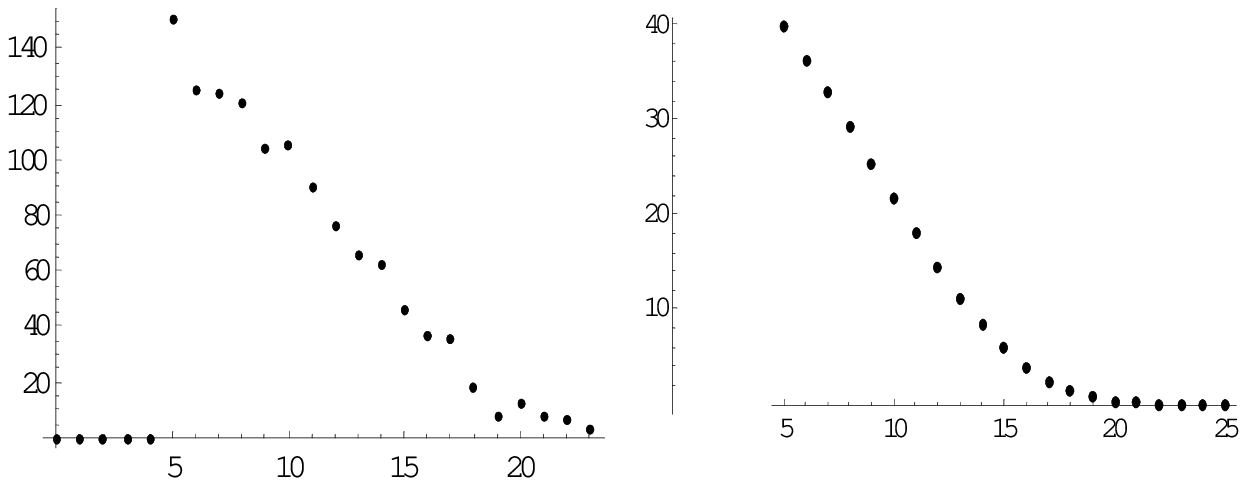}\\
Fig.8 Degree distributions of simulation for $n=1200$ and $k=5$ (left) and theory ($\gamma = 1/15$) (right) at $c=4.0$
\end{center}

In such a way, a critical point and behaviour of degree distributions before and after the critical point 
can be explained theoretically. 

\subsection{$A$-dependence}
We chose $A=30$  in order to analyze in a linear part of Eq.(1) in the presented model. 
When the value of $A$, however, changes, how results descibed in the previous sections differ? 
In this section we investigate the effects of $A$-value, especially small $A$. 

At $A=10$, the degree distribution function is such triangular in shape as Fig.9, which shows 
that the degree grows larger as  $k$ but drops at much larger $k$.  
This occurs because the probability $P$  rapidly decreases  at $A=10$ as $k$ grows larger. 
The  movement in the function  $P(k)$  at $A=10$ and $A=30$ when $c=4.0$ is shown in Fig.10.  
 When nodes with small $k$ first are connected to a newly joining node with greater probability in D-dominate world.   
So the population of nodes with a little larger $k$ increases. 
A similar phenomenon occurs every time newly joining nodes are addes to the network. 
These propagate into larger $k$ step by step.  
Nodes with excessively large $k$, however, are rarely connected by links to a newly joining nodes conversely, 
since the  probability $P$  rapidly decreases  as $k$ grows larger. 
The population of nodes with mean values of $k$ increses due to that reason, while nodes with larger $k$ 
are hard to increase.  

These facts show that degree distribution functions largely depend on the shape of a probability function. 

\begin{center}
\includegraphics[scale=0.8,clip]{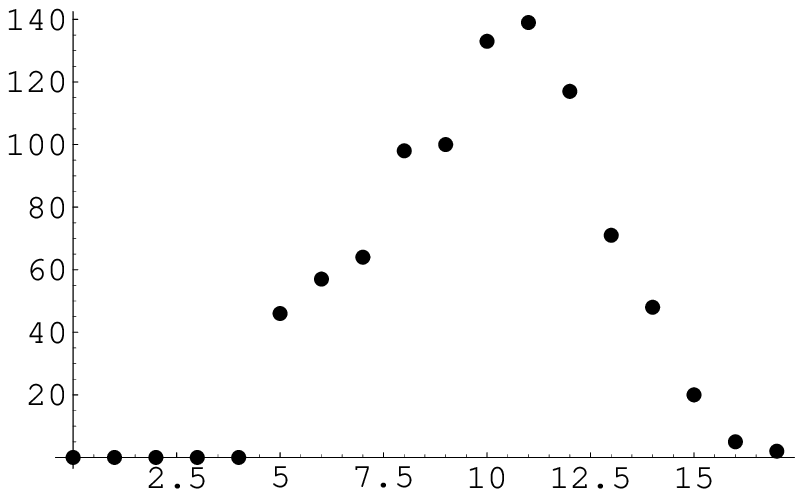}\\
Fig.9 Degree distributions for $n=900$, $k=5$  and  $c=4.0$ at $A=10$.

\includegraphics[scale=0.8,clip]{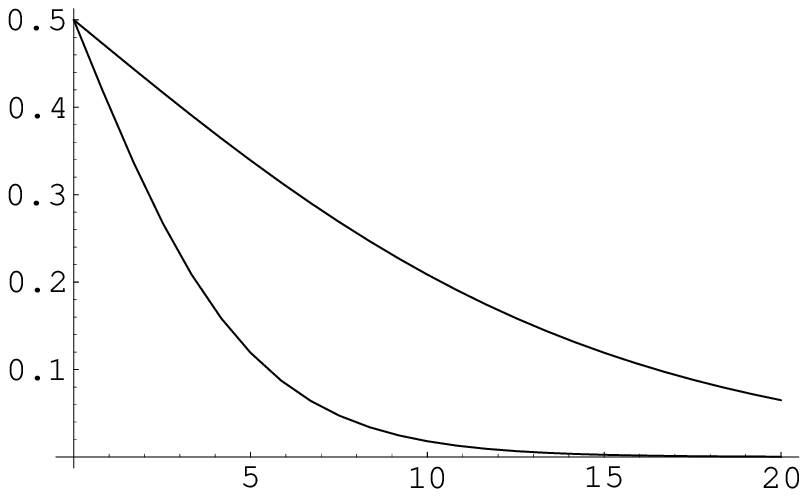}\\
Fig.10 $P(k)$ at $A=10$ (the lower curve) and $A=30$ (the upper curve) at $c=4.0$ 
\end{center}

\section{Summary}
In this article we proposed a model where game dynamics are working on a network and links are attached preferentially  
in proportion to total payoffs received by  nodes. 
According to the model, computer simulations are made to estimate degree distributions and analyse time series of strategies. 
We have found a sort of phase transition occurs with a order parameter  $c$. 
Though we only analyze a few phenomena in this article, but they can be  quantitatively explained 
in the theoretical point of view. 
The results are sensitive to the shape of a probability function that controls preferential attachment.  
Since there will be much interesting phenomena that should be explored in models 
that consider inner interactions among nodes in network growth, 
much wider properties of them should be investigated in details.

\end{document}